\documentclass[
reprint,
nofootinbib,
 amsmath,amssymb,
 aps,
pra,
twocolumn
]{revtex4-2}

\usepackage{graphicx}
\usepackage{dcolumn}
\usepackage{bm}
\usepackage{dsfont} 

\usepackage[english]{babel}

\usepackage[letterpaper,top=2cm,bottom=2cm,left=3cm,right=3cm,marginparwidth=1.75cm]{geometry}

\usepackage{amsmath}
\usepackage{graphicx}
\usepackage[colorlinks=true, allcolors=blue]{hyperref}

\usepackage{natbib}
\usepackage{ulem}
\usepackage{xcolor}
\usepackage{xspace}

\definecolor{jccol}{rgb}{0.2,0.8,0.2}

\newcommand{\VNe}{\ensuremath{V_{\mbox{\tiny Ne}}}\xspace}
\newcommand{\Vee}{\ensuremath{V_{\mbox{\tiny ee}}}\xspace}

\newcommand{\pur}{\ensuremath{P}\xspace}
\newcommand{\vis}{\ensuremath{\mathcal{V}}\xspace}
\newcommand{\viszer}{\ensuremath{\vis_0}\xspace}
\newcommand{\phizer}{\ensuremath{\phi_0}\xspace}
\newcommand{\visone}{\ensuremath{\vis_1}\xspace}
\newcommand{\phione}{\ensuremath{\phi_1}\xspace}
\newcommand{\rhoi}{\ensuremath{\rho_{\mbox{\tiny ion}}}\xspace}
\newcommand{\Tr}{\ensuremath{\mbox{tr}}\xspace}
\newcommand{\wR}{\ensuremath{\omega_{\mbox{\sc r}}}\xspace}
\newcommand{\wI}{\ensuremath{\omega_{\mbox{\sc i}}}\xspace}

\newcommand{\ket}[1]{\ensuremath{\left\vert#1\right\rangle}\xspace}

\begin{document}

\title{
Time-domain interferences as the source of electron-ion entanglement in Rabi-dressed photoemission
}
\author{L\'eonardo Rico}
\author{Morgan Berkane}
\author{Jonathan Dubois}
\author{J\'er\'emie Caillat}
\author{Richard Ta\"{\i}eb}
\author{Camille L\'ev\^eque}
\affiliation{Sorbonne Universit\'e, CNRS, 
Laboratoire de Chimie Physique-Mati\`ere et Rayonnement, LCPMR, F-75005 Paris, France}
\date{\today}

\begin{abstract}

We investigate bipartite entanglement between a photoelectron and its parent ion when the latter undergoes Rabi oscillations, following the recent experiment of [Nandi et al. \href{https://doi.org/10.1126/sciadv.ado0668} {Science Advances {\bf 10}, eado0668 (2024)}]. Using numerical simulations on a model atom, we show that this entanglement results from ionization events occurring at different times, with the photoelectron leaving the ion in distinct superpositions of internal states due to the Rabi coupling. Our interpretation brings forward the possibility to access the purity of the photoion state from photoelectron spectra. Furthermore, we demonstrate a tomographic reconstruction of the dressed ionic state dynamics from the observable spectra.

\end{abstract}

\maketitle

Entanglement lies at the heart of the second quantum revolution and has attracted increasing interest in multiple fields, including solid-state physics \cite{Simmons11,Takeda21}, cold atoms \cite{Kunkel18,Fadel18,Lange18,Yang20}, trapped ions \cite{Sackett00,Blinov04}, quantum information \cite{Almanakly25}, cryptography \cite{Yin20}, and more recently photoionization \cite{Vrakking21,Koll22}. The generation, preservation and detection of entanglement in quantum systems is now a central topic in the pursuit of quantum computing \cite{Graham22,Evered23}. At a more fundamental level, it is the cornerstone of the Einstein-Podolsky-Rosen paradox \cite{Einstein35}, in which a local measurement on one subsystem instantaneously influences the state of the other. 
In particular, entanglement between a photoelectron and its parent ion has recently gained considerable attention in the context of photoemission with coherent and intense light sources \cite{Vrakking21,Koll22,Ishikawa23,Laurell25,Berkane25,Shen25}. 

By essence, an atom that ionizes through a single channel does not exhibit entanglement between the photoelectron and the residual ion, as the total wavefunction in the final state factorizes into a single product of the two subsystems' states. 
In a recent experiment~\cite{Nandi24}, He atom interacting with a free-electron laser pulse \cite{McNeil10}
intense enough to simultaneously induce photoemission and Rabi oscillations in the parent ion was investigated. The appearance of a Grobe-Eberly (GE) doublet \cite{Grobe93} in the photoelectron spectrum was there identified as a clear manifestation of entanglement, in contrast with the photoemission process from a Rabi-dressed neutral atom leading to the so-called Autler–Townes doublet~\cite{Girju07,Rohringer08,Demekhin12_a,Demekhin12_b,Zhang14,Deng25,Wan25,Nandi22}. 
This dressed photoemission scheme has since then become a benchmark case for the study of light-induced entanglement between massive particles,
and it is the subject of recent  theoretical investigations~\cite{Ishikawa23,Stenquist25}.

\begin{figure*}[t]
\centering
\includegraphics[width=0.999\linewidth]{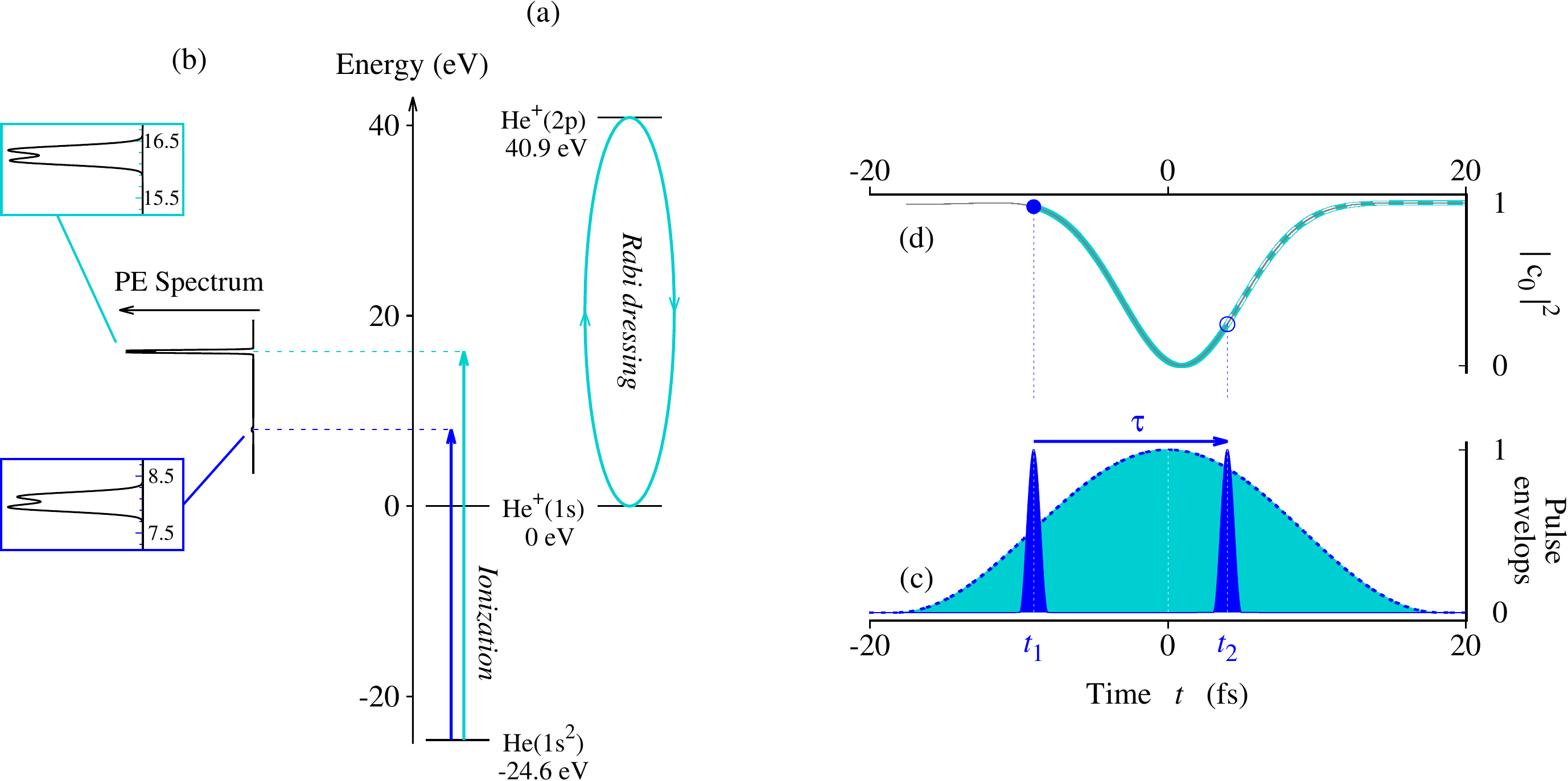}
\caption{\label{fig:sketch}  (a) Energy diagram  indicating the relevant He and He$^+$ levels  reproduced in our model atom. We considered a pulse of central frequency $\wR=40.9$ eV intense enough to simultaneously induce photoemission and resonant Rabi oscillations in the ion (light blue arrows) and a pulse of central frequency $\wI=32.7$ eV with a perturbative intensity inducing photoemission only (dark blue arrow). (b) Photoelectron spectrum obtained in presence of the two fields when both are assigned identical durations ($\approx 36$ fs). The peaks corresponding to the absorption of each photon are enlarged in the framed insets. They both display the typical GE doublet structure expected in Rabi-dressed photoemission. (c) Temporal profile of the pulse envelops considered in our pump-probe scheme: long dressing pulse at \wR (filled light blue), long ionizing pulse at \wI (dashed dark blue) and alternative composite short ionizing pulse at \wI (filled dark blue) (d) Interferometric pump-probe scheme involving two short ionizing pulses. The thin grey curve shows the population $\vert c_0(t) \vert ^2$  of field-free ionic ground state when the system evolves in the \ket{D_0(t)} dressed state (see text). A first ionizing pulse (here at time $t_1$) ``pumps'' the ion in the instantaneous dressed state $|\chi_{1}(t)\rangle$ (solid circle). The populated ionic state then evolves according to the light-blue curve. The second ionizing pulse (here at $t_2$) creates another photoelectron-photoion state (empty circle) which then interferes with the wave-packet created at $t_1$ (dashed curve). }
\end{figure*}

Here, we present a comprehensive mechanism for the generation of such entanglement and demonstrate how it enables experimental access to both the final purity and the tomographic reconstruction of the dressed ionic states dynamics. We base our study on numerical simulations with a model He atom, involving two active electrons in one dimension each, known in particular to quantitatively reproduce the GE doublet in this context \cite{Yu18}. Its Hamiltonian  expressed in atomic units (a.u.) reads
\begin{equation}\label{Hamiltonian}
H(t) = \sum_{i=1}^{2} \left\{\frac{[p_{i}+A(t)]^2}{2} + \VNe(x_{i})\right\} + \Vee(x_{1},x_{2}), \end{equation}
where the electron-nucleus (\VNe) and electron-electron (\Vee) interactions are modeled by soft-Coulomb potentials adjusted to reproduce the energies of ground state of He and of the $1s$ and $2p$ states of He\textsuperscript{+}~\cite{Yu18}, see Methods. The energy diagram of the essential states involved in the present work is given in Fig.~\ref{fig:sketch}(a). The interaction with the light pulse(s) in the various situations investigated in our study is implemented in the velocity gauge through the vector potential $A(t)$. We assigned to the latter  sine-squared temporal envelops, which are illustrated in Fig.~\ref{fig:sketch}(c). Total and channel-resolved photoelectron spectra were obtained from the numerical solution of the time-dependent Schr\"odinger equation using the tSurff method \cite{Tao12,Scrinzi12}.  For the analysis and interpretation of our simulations, we will resort to two different ionic state basis. The first one  comprises the field-free eigenstates of the ion, i.e. He$^+(1s)$ and He$^+(2p)$, denoted $\ket{g}$ and  $\ket{e}$ respectively. 
The second one, which evolves in time, is built by backpropagating the field free basis in presence of the dressing field. The corresponding states are noted \ket{D_0} and \ket{D_1}, converging at $t\rightarrow \infty$ toward $\ket{g}$ and  $\ket{e}$, respectively.(See Methods).

\begin{figure}[t]
\centering
\includegraphics[width=0.95\linewidth]{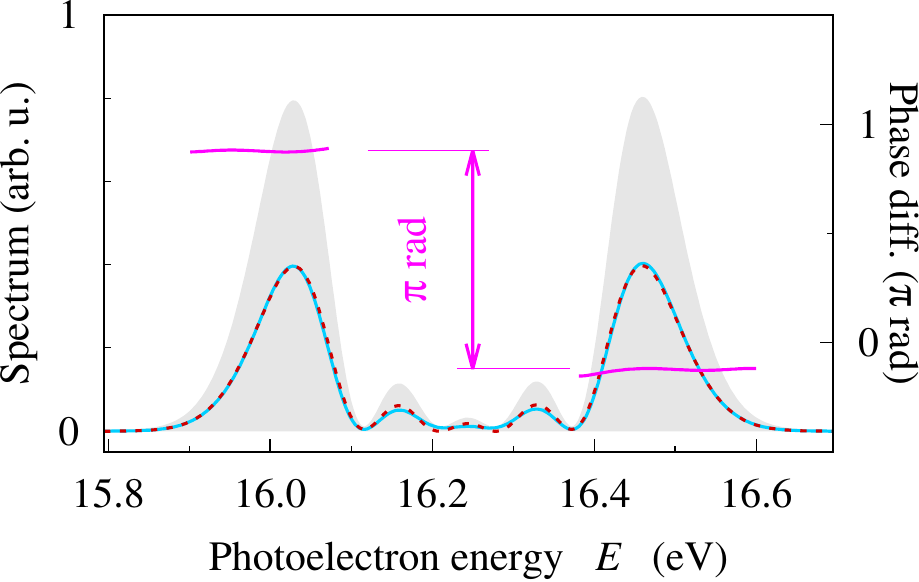}
\caption{\label{fig:spectrum} Total (shaded gray area) and channel-resolved [solid cyan line He$^+(1s)$; red dashed line He$^+(2p)$] photoelectron spectra for $I_{R}=5.7\times10^{13}$ W.cm$^{-2}$, $\omega_{R}=40.86$ eV and $T_{R}=36.43$ fs. The relative phase experienced a jump of $\pi$-rad from the first peak ($\sim 16.0$ eV) to the second one ($\sim 16.5$ eV), indicated by the pink lines. }
\end{figure}

We first addressed the benchmark situation investigated in Ref.~\cite{Yu18}. For this, we considered a single pulse intense enough to simultaneously ionize the atom and excite the ion. Its central frequency was set to $\omega_{R}=40.86$ eV which resonantly triggers Rabi oscillations between the ionic ground and first excited states with a frequency $\Omega=2\mathcal{E}_{0}d_{ge}$, where $d_{ge}$ is the transition dipole moment and $\mathcal{E}_0$ the field peak strength. The transitions involved in the process are indicated with light-blue arrows in Fig.~\ref{fig:sketch}(a). We used a total pulse duration of $T_{R}=360$ laser cycles ($\sim 36.43$ fs) and an intensity of $I_R=5.6\times10^{13}$ W.cm$^{-2}$, which spans $\sim 2.5$ Rabi periods and ensures the appearance of the GE doublet in the photoelectron peak, see Fig.~\ref{fig:spectrum}. Its features are identical to the one shown in~\cite{Yu18}. To quantify the photoelectron-photoion entanglement evidenced by the GE doublet, we computed the purity $\pur=\Tr(\rhoi^2)/\Tr(\rhoi)^2$ out of the ion's reduced density matrix (RDM) $\rhoi$ built on the final wave-function, see Methods.  In line with Ref. \cite{Nandi24}, we find $\pur=0.50$, indicating a maximal entanglement between the photoemission products. At this point, one clearly sees that the photoelectron {\it spectrum} alone displays no evidence of this entanglement, since the channel-resolved peak structures are nearly identical in ground- and excited-state channels, see Fig.~\ref{fig:spectrum}. However, we found that the relative phase between the two channels (pink line in Fig.~\ref{fig:spectrum}) undergoes a $\pi$-rad spectral jump between the two peaks of the doublet. This prevents the factorization of the final state wave function and thus entangles the system. This strongly suggests a time-domain interpretation of the investigated entanglement build-up.

In the field configuration considered so far, it is tedious to establish the underlying physical mechanism, since the same light pulse induces both the ionization process and Rabi oscillations in the ion~\cite{Grobe93, Nandi24}. To overcome this difficulty, we performed a series of additional simulations with purposely designed pulses.
We used a two-color scheme to unambiguously discriminate the Rabi dressing from the photoemission process. While keeping the same pulse as before to drive the Rabi oscillations, we added time-shifted ionizing pulses at frequency $\omega_{\text{I}}=32.65$ eV adjusted to induce a photoelectron peak well separated from the GE doublet, see Fig. \ref{fig:sketch}(c,b). 

First, when both the ionizing and Rabi pulses have the same duration, the GE doublet is imprinted in the low energy part of the photo electron spectrum, $E\sim 8.1$eV, resulting from the absorption of one $\omega_I$ photon, see Fig. \ref{fig:sketch}(b). This shows that the signature of the Rabi-dressing of the ionic states is independent of the ionization process. 

Then, we use now a pair of ionizing pulses with an envelop lasting  $1.90$ fs (15 cycles), i.e., much shorter than the Rabi period. This ensures to achieve a satisfactory time resolution. In the following, we denote $t_1$ and $t_2$ the central times of the ionizing pulse envelops and $\tau=t_2-t_1$ their tunable relative delay, see Fig.~\ref{fig:sketch} (c). Without loss of generality, we set $t_1$ at the maximum of the Rabi pulse, i.e., $t_1=0$. This field sequence is analogous to one used for Ramsey-like interferometry \cite{Ramsey50}.

\begin{figure*}[t]
\centering
\includegraphics[width=0.999999999\textwidth]{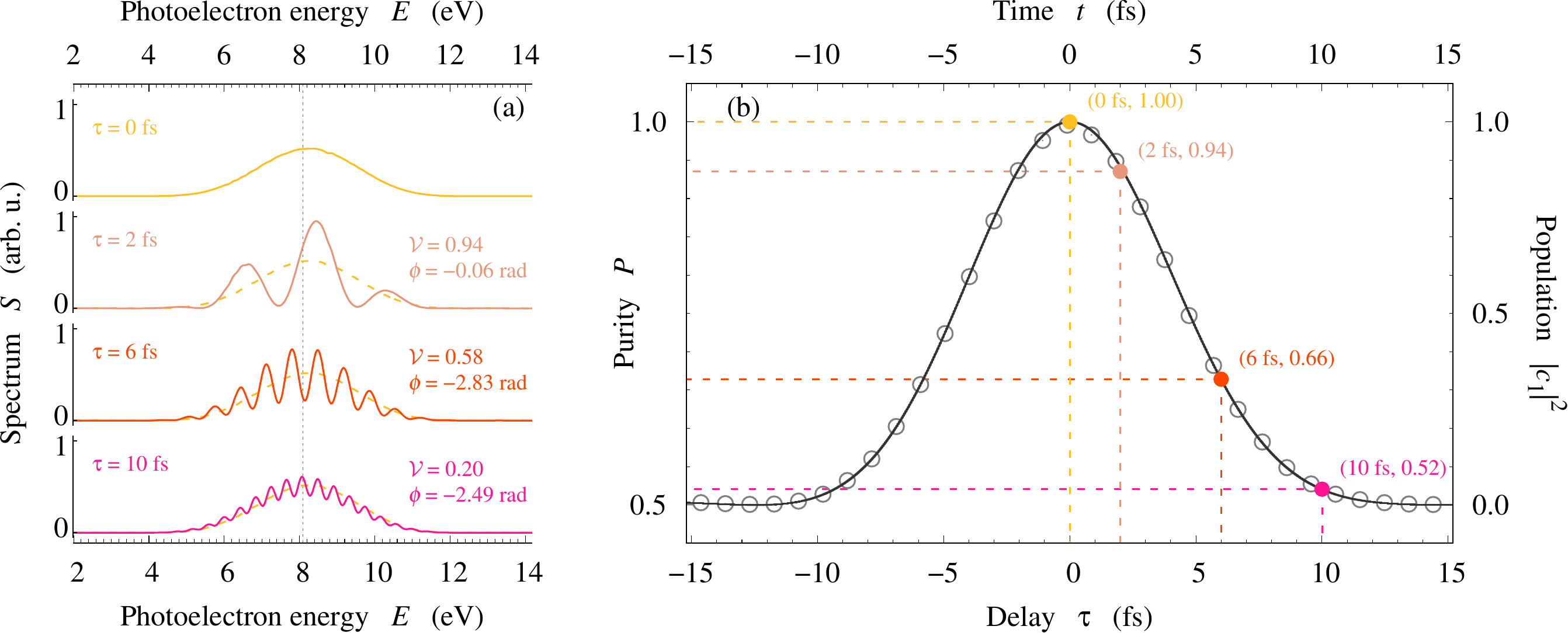}
\caption{\label{Fig3} (a) Photoelectron spectra for four illustrative delays $\tau=0$ fs, $2$ fs, $6$ fs and $10$ fs (top-down).  The single-pulse case is reproduced in dashed line as a guide for all non-vanishing $\tau$ curves, where the values of the  visibility \vis and phase $\phi$, obtained by fitting Eq.~\eqref{tot_spec} to the numerical spectra,  are indicated.
 The vertical dotted line indicate the average energy ($8.08$ eV) in the $\tau=0$ case. (b) Population $|c_1|^2=\vert\langle  D_1(t) \vert g \rangle\vert^2$  as a function of time  $t$ (black line). The circles correspond to the values of the purity from the TDSE simulations and the colored bullets to the ones obtained from $\mathcal{V}$ [Eq.~\eqref{P_C}] in the spectra of panel (a). The ionizing pulses are defined with $\omega_I=32.65$ eV, an intensity $I_{I}=3.5\times10^{13}$ W.cm$^{-2}$ and a total pulse duration $T_I=1.90$ fs. Note that the characteristics of the pulse driving the Rabi oscillations are identical to the one of Fig. \ref{fig:spectrum}, except for a reduced intensity ($I_R=1.1\times10^{13}$~W.cm$^{-2}$) to span a single Rabi period within the \wR pulse duration.  }
\end{figure*}

Figure \ref{Fig3} displays photoelectron spectra for three illustrative ionization delays ($\tau=0, 2, 6$ and $10$ fs), along with the corresponding final purity values. At $\tau=0$, the two pulses effectively merge, resulting in $P=1$, i.e., no entanglement. This shows that Rabi oscillations alone, when ionization occurs at a well-defined instant, do not generate entanglement. In other words, the dynamics in the ion does not affect the photoelectron state upon ionization. In contrast, when the delay increases, spectral oscillations appear, with a period of $2\pi/\tau$, reminiscent of Ramsay fringes. They are accompanied by a decrease of \pur, reaching a minimum at half a Rabi period.  

This variation of the final purity with the delay provides valuable physical insight into the entanglement mechanism. As $\tau$ increases, ionization occurs from increasingly different superpositions of Rabi-dressed states. Entanglement then arises from spectral interferences between the two photoelectron wavepackets created at times $t_1$ and $t_2$. This is similar to the mechanism first designed and investigated in \cite{Vrakking21}. Panel (b) of Fig.~\ref{Fig3} shows a clear correlation between the purity evolution (black stars) and the population of the field-free ionic ground state (orange line) as a function of $\tau$. 

To formalize how this entanglement is imprinted in the observable photoelectron spectra, we now establish an explicit relationship between the visibility of the $\tau$-dependent oscillations and the corresponding value of \pur.
 To this end, we express the normalized wavefunction of the final ionized system as the coherent superposition 
\begin{equation}\label{tot_wf}
  |\Psi(t)\rangle=\frac{1}{\sqrt{2}}\sum_{i=1}^{2}|\chi_i(t)\rangle\otimes|\psi_{i}(t)\rangle
\end{equation}
of two pure product-states of the electron-ion pair created at $t_1$ and $t_2$ respectively 
In each contribution ($i=1,2$), $|\psi_{i}(t)\rangle$ represents the photoelectron wave packet, while $|\chi_i(t)\rangle$ represents the ionic state launched within the Rabi pulse with the initial condition $|\chi_i(t_{i})\rangle=|g\rangle$. Each photoelectron wavepacket undergoes a field free evolution,
\begin{equation}
|\psi_{i}(t)\rangle=\int\, \psi_i(E)\vert E\rangle e^{-iEt}\, dE, 
\end{equation}
where $E$ is the photoelectron energy referred to the ionization threshold and $\vert E \rangle$ the continuum eigenstates. The associated spectral amplitudes are identical up to a phase difference accumulated between the respective birth times, i.e.,
\begin{equation}
\psi_{2}(E)=\psi_{1}(E)e^{iE\tau}.
\end{equation}
For the sake of convenience, we switch to the Rabi-dressed basis and express the evolution of each ionic state as
\begin{equation}
  |\chi_{i}(t)\rangle=\sum_{j=0}^{1}c_j(t_i)|D_{j}(t)\rangle, 
\end{equation}
In this  framework, the state-resolved photoelectron spectra read
\begin{equation}
S_{g,e}(E)=\frac{1}{2} |\psi_1(E)|^2 \left| c_{0,1}(t_1)+c_{0,1}(t_2)e^{iE\tau} \right|^2,
\end{equation} 
and the total spectrum is then expressed as 
\begin{equation}\label{tot_spec}
S(E)=  |\psi_1(E)|^2\left[1+\vis\times\cos{(E\tau+\phi)}\right].
\end{equation}
Although it corresponds to the incoherent sum $S_{g}(E)+S_{e}(E)$ on the channel basis, it bears signatures of interferences that appear when considering the dressed-state basis. They materialize as the $\tau$-dependent  visibility and phase of the oscillatory patterns in the spectra, respectively
\begin{equation}\label{phase_contr}
\begin{split}
\vis&=\left|\sum_{j=0}^{1}c_j(t_1)c^{*}_j(t_2)\right|\\
\phi&=\arg\left\{\sum_{j=0}^{1}c_j(t_1)c^{*}_j(t_2)\right\}.
\end{split}
\end{equation}
To obtain an explicit relationship between \vis and the purity \pur, we consider the photoion's RDM $\rho_{ion}$ in the final state (i.e. $t>T_R$ and we discard the $t$-dependencies in the following derivations). 
 Since time-dependent  ionic states are not necessarily orthogonal, we introduce the projection $|\chi_{\perp}\rangle=(1-\vert \chi_1 \rangle\langle \chi_1\vert)\vert \chi_2 \rangle$ and the overlaps $\alpha=\langle \chi_1|\chi_2\rangle$ and $\beta=\langle \chi_{\perp}|\chi_2\rangle$. Expanding the total final wavefunction $|\Psi\rangle$ (Eq. \ref{tot_wf}) on the orthogonal $\{\vert \chi_1\rangle,\vert \chi_\perp\rangle\}$ basis and tracing out the photoelectron degree of freedom in the total density matrix leads to the compact expression   
\begin{equation}
\rho_{ion}=\frac{1}{2}
\begin{pmatrix}
  1+|\alpha|^2 & \alpha^{*}\beta\\ 
  \alpha\beta^{*} & |\beta|^2
\end{pmatrix}.
\end{equation}
By furthermore noting that the visibility can be expressed as  $\vis=|\langle\chi_1|\chi_2\rangle|$ at all times, such that $|\alpha|^2=\vis^2$ and $|\beta|^2=1-\vis^2$, we reach the simple and direct formal mapping,
\begin{equation}\label{P_C}
P=\frac{1}{2}(1+\vis^2).
\end{equation}
This opens a promising avenue to experimentally obtain the purity of the final bipartite system from the visibility of the oscillations in the measurable photoelectron spectrum. An identical mapping was established in \cite{Dittel21}, in the context of matter-wave interference when the particle entangles with the quantum state of the double slits. 

The validity of this result is assessed in Fig.~\ref{Fig3}. We find a perfect agreement between the numerically exact purity extracted from the simulations (open circles) and the ones extracted directly from the visibility (colored circles) of the photoelectron spectra displayed in panel (a). Moreover, at $\tau=0$, $t_1=t_2$ and thus $\vis=|\langle\chi_1|\chi_1\rangle|=1$, leading to a purity $\pur=1$, i.e., no entanglement.   
Finally, we have checked that, when using many ionizing pulses sampling the whole Rabi pulse, we recover the GE doublet. This validates our interpretation of interferences in the time domain as a source of entanglement.

Actually, Eq. \ref{tot_spec} contains significantly more physically valuable information, opening the way to a complete tomography of the ionic dynamics in the dressed state basis. 
Indeed, the time evolution of $c_{0}$ is conveniently obtained by setting one ionizing pulse at $t_1>T_R$, such that the ionic states are not dressed, notably $|D_0(t_1)\rangle=|g\rangle$, thus $c_0(t_1)=1$ and $c_1(t_1)=0$. By noting \viszer and \phizer the visibility and phase in this case, indentifying $t$ and $t_2$, Eq.~\ref{phase_contr} reduces to
\begin{equation}\label{eqn:c0}
c_0(t)=\viszer e^{-i\phizer}.
\end{equation}
Thus, we directly access the modulus and phase of $c_{0}(t)$ from the measured spectra.  

Then, the second coefficient, $c_{1}(t)$, is obtained by setting $t_1$ to any other value inside the Rabi pulse and measuring the visibility and phase (\visone and \phione), while scanning $t_2$. Using again Eq.~\ref{phase_contr} and the previously reconstructed $c_{0}(t)$, we obtain for each $t=t_2$,
\begin{equation}\label{eqn:c1}
c_1(t)=\mathcal{N}\left({\visone}e^{-i{\phione}}-c_0^\star(t_1)c_0(t)\right).
\end{equation}
The modulus of the reference-dependent coefficient $\mathcal{N}$ is equal to $1/\sqrt{1-\vert c_0(t_1) \vert^2}$ and its phase depends on an arbitrary origin. In the results presented hereafter, we simply set $\arg(\mathcal{N})=0$. Note that, for the resonant case, a convenient choice for $t_1$ is at the maximum of population inversion, which leads to $c_0(t_1)=0$ and cancels the last term in Eq.~\eqref{eqn:c1}.

\begin{figure}[t]
\centering
\includegraphics[width=0.9\linewidth]{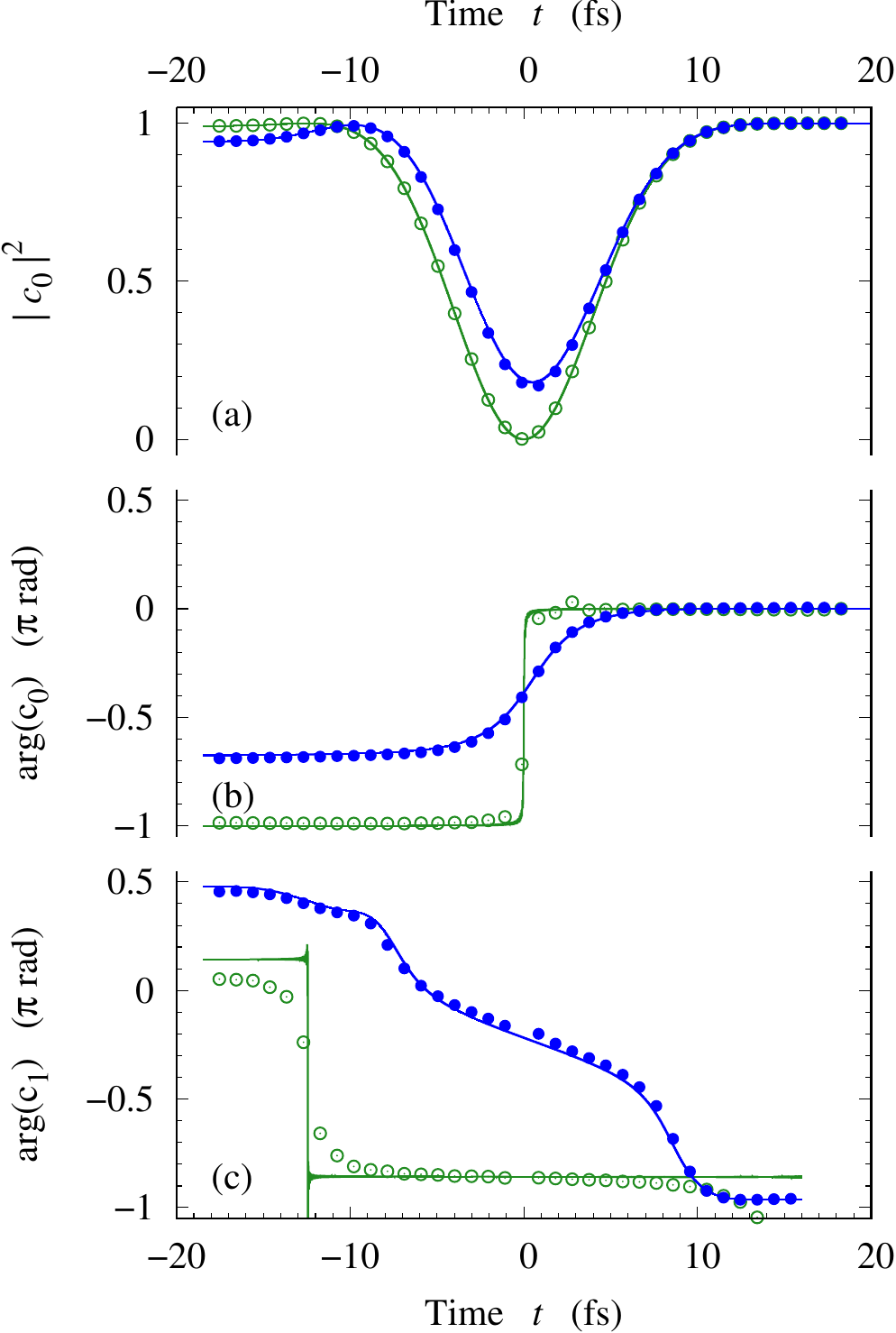}
\caption{\label{Fig4}
Ion dynamics in the dressed states: tomographic reconstruction (symbols) and exact values (solid lines). (a): Population $\vert c_0(t)\vert^2=\vert\langle D_0(t)|g\rangle\vert^2$ as a function of time for $\Delta=0$~meV (green) and $\Delta=75.34$~meV (blue). (b) and (c): Phases of $c_0(t)$ and $c_1(t)=\langle D_1(t)|g\rangle$, respectively. Note that we only show arg$(c_1)$ when the population is significant, i.e., $|c_1|^2\gtrsim10^{-3}$.}
\end{figure}

To illustrate our procedure, we considered the resonant case discussed so far, i.e., $\Delta=\omega_R-(E_e-E_g)=0$, as well as a detuning of $\Delta=75.34$ meV which leads to a population inversion of $\sim80\%$. As seen in Fig. \ref{Fig4} (a-b), the reconstructed coefficient $c_{0}(t)$ (symbols) is in perfect agreement with the numerically exact ion dynamics (solid line), for both $\Delta=0$~meV (green/circles) and $\Delta=75.34$~meV (blue/bullets).   
The phase of $c_1(t)=\langle D_1(t)|g\rangle$ extracted from the tomography  is displayed Fig. \ref{Fig4}(c), and is also in perfect agreement with the simulation  for the resonant  and non-resonant case [same color code as in frame (a)].

In this Letter, we presented an experimentally feasible scheme for quantum state tomography of Rabi-dressed states using only the photoelectron spectra. This scheme is quite general and not limited to Rabi oscillations in ionic states of atoms, as it requires only a few conditions: (i) synchronized and delay-controlled ionizing pulses that are short enough such that $T_I \ll 2\pi/\Omega$, and (ii) an accessible transition for the dressing within a high-intensity laser pulse. Our work enhances the understanding of bipartite entanglement generation in the context of photoionization. We show that entanglement arises from ionization at different times, which leaves the ionic dressed parent in a different state superposition. The entanglement appears in the momentum imaging of the photoelectrons.      

 The authors wish to acknowledge fruitful and stimulating numerous exchanges with Charles Bourassin-Bouchet and Pascal Salières. This research received financial support from the French National Research Agency through grants No. ANR-20-CE30-0007-DECAP, ANR-24-CE29-0141-EPAD, and No. ANR-21-190
CE30-0036-03-ATTOCOM. JD acknowledges funding from the European Union’s Horizon
Europe research and innovation program under the Marie Sklodowska-Curie Grant Agreement No. 101154681. Views and opinions expressed are, however, those of the author(s)
only and do not necessarily reflect those of the European Union or the European Commission. Neither the European Union nor the European Commission can be held responsible
for them.

\section*{Methods}
\textbf{Numerical simulations} were carried out 
on a $2\otimes$1D-electron model atom based on the hamiltonian given in Eq.~\eqref{Hamiltonian} with the potentials
\begin{equation}
\begin{split}    
V_{N}(x_{i})&=-\frac{1.1225}{\sqrt{x_{i}^2+a^2}}\\
V_{e}(x_{1},x_{2})&=\frac{0.6317}{\sqrt{(x_{1}-x_{2})^2+a^2}}
\end{split}
\end{equation}
with $a=0.3028$ a.u., following Ref. \cite{Yu18}. The resulting field-free energies are $E(1s^2)=-2.9057$~a.u., $E(1s)=-2.0014$~a.u. and $E(2p)=-0.4999$~a.u., for the He ground-state and for the He$^+$ first and second excited states, respectively. The interaction was implemented in the velocity gauge with generic $\sin^2$ envelops of variable total duration $T$
\begin{eqnarray}
f(t)=\cos^2\left(\frac{\pi}{T}t\right)\mathds{1}_{t\in[-\frac{T}{2},+\frac{T}{2}]},
\end{eqnarray}
where $\mathds{1}$ is the indicator function, on the vector potential. The time evolution of the wavefunction were obtained by numerically solving the time-dependent Schr\"odinger equation using the $BM_64$ algorithm \cite{Blanes02,McLachlan22}, with a time step of $6.728\times 10^{-3}$ a.u. and a spatial step of $\Delta x=0.25$ a.u. with 1024 points.\\
\textbf{Photoelectron spectra} were obtained using the time dependent surface flux method (t-SURFF)~\cite{Scrinzi12}, with a radius $R_{c}=50$~a.u. beyond which the wavefunction is propagated analytically, along the coordinate of each electron. Because the ionic states are dressed, the electronic ionized wavefunction is projected onto the time-dependent ionic states at the flux surface \cite{Tao12}. These states are obtained by propagating backward the time-dependent Schr\"odinger equation for He$^{+}$ and are denoted $|D_{i}(t)\rangle$ in the main text.  
\newline\textbf{Absorbing Boundary conditions} are used to keep the 2D grid small and avoid nonphysical reflection, a $\cos^{1/8}$ mask-function acting over $70$ a.u. on the edges of the simulation grid. 
\\
\textbf{The purity} of the final system is obtained from the ionic reduced density matrix, $\rho_{ion}=\Tr_{e}(\rho)$, where
\begin{equation}
\rho=\lim_{T \rightarrow \infty}|\Psi(T)\rangle\langle\Psi(T)|
\end{equation}
is the total density matrix when the electric fields are off, and $\Tr_{e}$ indicates the trace over the photoelectron degree of freedom. The normalized purity reads then, $\pur=\Tr(\rhoi^2)/\Tr(\rhoi)^2$. In our case, since the system is bipartite, the purity obtained from the ionic RDM or the photoelectron RDM are identical, and we chose the former for convenience.\\  

\bibliographystyle{apsrev4-1}
\bibliography{biblio}

\begin{thebibliography}{38}%
\makeatletter
\providecommand \@ifxundefined [1]{%
 \@ifx{#1\undefined}
}%
\providecommand \@ifnum [1]{%
 \ifnum #1\expandafter \@firstoftwo
 \else \expandafter \@secondoftwo
 \fi
}%
\providecommand \@ifx [1]{%
 \ifx #1\expandafter \@firstoftwo
 \else \expandafter \@secondoftwo
 \fi
}%
\providecommand \natexlab [1]{#1}%
\providecommand \enquote  [1]{``#1''}%
\providecommand \bibnamefont  [1]{#1}%
\providecommand \bibfnamefont [1]{#1}%
\providecommand \citenamefont [1]{#1}%
\providecommand \href@noop [0]{\@secondoftwo}%
\providecommand \href [0]{\begingroup \@sanitize@url \@href}%
\providecommand \@href[1]{\@@startlink{#1}\@@href}%
\providecommand \@@href[1]{\endgroup#1\@@endlink}%
\providecommand \@sanitize@url [0]{\catcode `\\12\catcode `\$12\catcode `\&12\catcode `\#12\catcode `\^12\catcode `\_12\catcode `\%12\relax}%
\providecommand \@@startlink[1]{}%
\providecommand \@@endlink[0]{}%
\providecommand \url  [0]{\begingroup\@sanitize@url \@url }%
\providecommand \@url [1]{\endgroup\@href {#1}{\urlprefix }}%
\providecommand \urlprefix  [0]{URL }%
\providecommand \Eprint [0]{\href }%
\providecommand \doibase [0]{http://dx.doi.org/}%
\providecommand \selectlanguage [0]{\@gobble}%
\providecommand \bibinfo  [0]{\@secondoftwo}%
\providecommand \bibfield  [0]{\@secondoftwo}%
\providecommand \translation [1]{[#1]}%
\providecommand \BibitemOpen [0]{}%
\providecommand \bibitemStop [0]{}%
\providecommand \bibitemNoStop [0]{.\EOS\space}%
\providecommand \EOS [0]{\spacefactor3000\relax}%
\providecommand \BibitemShut  [1]{\csname bibitem#1\endcsname}%
\let\auto@bib@innerbib\@empty
\bibitem [{\citenamefont {Simmons}\ \emph {et~al.}(2011)\citenamefont {Simmons}, \citenamefont {Brown}, \citenamefont {Riemann} \emph {et~al.}}]{Simmons11}%
  \BibitemOpen
  \bibfield  {author} {\bibinfo {author} {\bibfnamefont {S.}~\bibnamefont {Simmons}}, \bibinfo {author} {\bibfnamefont {R.}~\bibnamefont {Brown}}, \bibinfo {author} {\bibfnamefont {H.}~\bibnamefont {Riemann}},  \emph {et~al.},\ }\href {\doibase 10.1038/nature09696} {\bibfield  {journal} {\bibinfo  {journal} {Nature}\ }\textbf {\bibinfo {volume} {470}},\ \bibinfo {pages} {69–72} (\bibinfo {year} {2011})}\BibitemShut {NoStop}%
\bibitem [{\citenamefont {Takeda}\ \emph {et~al.}(2021)\citenamefont {Takeda}, \citenamefont {Noiri}, \citenamefont {Nakajima} \emph {et~al.}}]{Takeda21}%
  \BibitemOpen
  \bibfield  {author} {\bibinfo {author} {\bibfnamefont {K.}~\bibnamefont {Takeda}}, \bibinfo {author} {\bibfnamefont {A.}~\bibnamefont {Noiri}}, \bibinfo {author} {\bibfnamefont {T.}~\bibnamefont {Nakajima}},  \emph {et~al.},\ }\href {\doibase 10.1038/s41565-021-00925-0} {\bibfield  {journal} {\bibinfo  {journal} {Nat. Nanotechnol.}\ }\textbf {\bibinfo {volume} {16}},\ \bibinfo {pages} {965–969} (\bibinfo {year} {2021})}\BibitemShut {NoStop}%
\bibitem [{\citenamefont {Kunkel}\ \emph {et~al.}(2018)\citenamefont {Kunkel} \emph {et~al.}}]{Kunkel18}%
  \BibitemOpen
  \bibfield  {author} {\bibinfo {author} {\bibfnamefont {P.}~\bibnamefont {Kunkel}} \emph {et~al.},\ }\href {\doibase 10.1126/science.aao2254} {\bibfield  {journal} {\bibinfo  {journal} {Science}\ }\textbf {\bibinfo {volume} {360}},\ \bibinfo {pages} {413} (\bibinfo {year} {2018})}\BibitemShut {NoStop}%
\bibitem [{\citenamefont {Fadel}\ \emph {et~al.}(2018)\citenamefont {Fadel} \emph {et~al.}}]{Fadel18}%
  \BibitemOpen
  \bibfield  {author} {\bibinfo {author} {\bibfnamefont {M.}~\bibnamefont {Fadel}} \emph {et~al.},\ }\href {\doibase 10.1126/science.aao1850} {\bibfield  {journal} {\bibinfo  {journal} {Science}\ }\textbf {\bibinfo {volume} {360}},\ \bibinfo {pages} {409} (\bibinfo {year} {2018})}\BibitemShut {NoStop}%
\bibitem [{\citenamefont {Lange}\ \emph {et~al.}(2018)\citenamefont {Lange} \emph {et~al.}}]{Lange18}%
  \BibitemOpen
  \bibfield  {author} {\bibinfo {author} {\bibfnamefont {K.}~\bibnamefont {Lange}} \emph {et~al.},\ }\href {\doibase 10.1126/science.aao2035} {\bibfield  {journal} {\bibinfo  {journal} {Science}\ }\textbf {\bibinfo {volume} {360}},\ \bibinfo {pages} {416} (\bibinfo {year} {2018})}\BibitemShut {NoStop}%
\bibitem [{\citenamefont {Yang}\ \emph {et~al.}(2020)\citenamefont {Yang} \emph {et~al.}}]{Yang20}%
  \BibitemOpen
  \bibfield  {author} {\bibinfo {author} {\bibfnamefont {B.}~\bibnamefont {Yang}} \emph {et~al.},\ }\href {\doibase 10.1126/science.aaz6801} {\bibfield  {journal} {\bibinfo  {journal} {Science}\ }\textbf {\bibinfo {volume} {369}},\ \bibinfo {pages} {550} (\bibinfo {year} {2020})}\BibitemShut {NoStop}%
\bibitem [{\citenamefont {Sackett}\ \emph {et~al.}(2000)\citenamefont {Sackett}, \citenamefont {Kielpinski}, \citenamefont {King}, \citenamefont {Langer}, \citenamefont {Meyer} \emph {et~al.}}]{Sackett00}%
  \BibitemOpen
  \bibfield  {author} {\bibinfo {author} {\bibfnamefont {C.~A.}\ \bibnamefont {Sackett}}, \bibinfo {author} {\bibfnamefont {D.}~\bibnamefont {Kielpinski}}, \bibinfo {author} {\bibfnamefont {B.~E.}\ \bibnamefont {King}}, \bibinfo {author} {\bibfnamefont {C.}~\bibnamefont {Langer}}, \bibinfo {author} {\bibfnamefont {V.}~\bibnamefont {Meyer}},  \emph {et~al.},\ }\href {\doibase 10.1038/35005011} {\bibfield  {journal} {\bibinfo  {journal} {Nature}\ }\textbf {\bibinfo {volume} {404}},\ \bibinfo {pages} {256–259} (\bibinfo {year} {2000})}\BibitemShut {NoStop}%
\bibitem [{\citenamefont {Blinov}\ \emph {et~al.}(2004)\citenamefont {Blinov}, \citenamefont {Moehring}, \citenamefont {Duan} \emph {et~al.}}]{Blinov04}%
  \BibitemOpen
  \bibfield  {author} {\bibinfo {author} {\bibfnamefont {B.}~\bibnamefont {Blinov}}, \bibinfo {author} {\bibfnamefont {D.}~\bibnamefont {Moehring}}, \bibinfo {author} {\bibfnamefont {L.~.}\ \bibnamefont {Duan}},  \emph {et~al.},\ }\href {\doibase 10.1038/nature02377} {\bibfield  {journal} {\bibinfo  {journal} {Nature}\ }\textbf {\bibinfo {volume} {428}},\ \bibinfo {pages} {153–157} (\bibinfo {year} {2004})}\BibitemShut {NoStop}%
\bibitem [{\citenamefont {Almanakly}\ \emph {et~al.}(2025)\citenamefont {Almanakly}, \citenamefont {Yankelevich}, \citenamefont {Hays} \emph {et~al.}}]{Almanakly25}%
  \BibitemOpen
  \bibfield  {author} {\bibinfo {author} {\bibfnamefont {A.}~\bibnamefont {Almanakly}}, \bibinfo {author} {\bibfnamefont {B.}~\bibnamefont {Yankelevich}}, \bibinfo {author} {\bibfnamefont {M.}~\bibnamefont {Hays}},  \emph {et~al.},\ }\href {\doibase 10.1038/s41567-025-02811-1} {\bibfield  {journal} {\bibinfo  {journal} {Nat. Phys.}\ }\textbf {\bibinfo {volume} {21}},\ \bibinfo {pages} {825–830} (\bibinfo {year} {2025})}\BibitemShut {NoStop}%
\bibitem [{\citenamefont {Yin}\ \emph {et~al.}(2020)\citenamefont {Yin}, \citenamefont {Li}, \citenamefont {Liao} \emph {et~al.}}]{Yin20}%
  \BibitemOpen
  \bibfield  {author} {\bibinfo {author} {\bibfnamefont {J.}~\bibnamefont {Yin}}, \bibinfo {author} {\bibfnamefont {Y.-H.}\ \bibnamefont {Li}}, \bibinfo {author} {\bibfnamefont {S.-K.}\ \bibnamefont {Liao}},  \emph {et~al.},\ }\href {\doibase 10.1038/s41586-020-2401-y} {\bibfield  {journal} {\bibinfo  {journal} {Nature}\ }\textbf {\bibinfo {volume} {582}},\ \bibinfo {pages} {501–505} (\bibinfo {year} {2020})}\BibitemShut {NoStop}%
\bibitem [{\citenamefont {Vrakking}(2021)}]{Vrakking21}%
  \BibitemOpen
  \bibfield  {author} {\bibinfo {author} {\bibfnamefont {M.~J.~J.}\ \bibnamefont {Vrakking}},\ }\href {\doibase 10.1103/PhysRevLett.126.113203} {\bibfield  {journal} {\bibinfo  {journal} {Phys. Rev. Lett.}\ }\textbf {\bibinfo {volume} {126}},\ \bibinfo {pages} {113203} (\bibinfo {year} {2021})}\BibitemShut {NoStop}%
\bibitem [{\citenamefont {Koll}\ \emph {et~al.}(2022)\citenamefont {Koll}, \citenamefont {Maikowski}, \citenamefont {Drescher}, \citenamefont {Witting},\ and\ \citenamefont {Vrakking}}]{Koll22}%
  \BibitemOpen
  \bibfield  {author} {\bibinfo {author} {\bibfnamefont {L.-M.}\ \bibnamefont {Koll}}, \bibinfo {author} {\bibfnamefont {L.}~\bibnamefont {Maikowski}}, \bibinfo {author} {\bibfnamefont {L.}~\bibnamefont {Drescher}}, \bibinfo {author} {\bibfnamefont {T.}~\bibnamefont {Witting}}, \ and\ \bibinfo {author} {\bibfnamefont {M.~J.~J.}\ \bibnamefont {Vrakking}},\ }\href {\doibase 10.1103/PhysRevLett.128.043201} {\bibfield  {journal} {\bibinfo  {journal} {Phys. Rev. Lett.}\ }\textbf {\bibinfo {volume} {128}},\ \bibinfo {pages} {043201} (\bibinfo {year} {2022})}\BibitemShut {NoStop}%
\bibitem [{\citenamefont {Graham}\ \emph {et~al.}(2022)\citenamefont {Graham}, \citenamefont {Song}, \citenamefont {Scott} \emph {et~al.}}]{Graham22}%
  \BibitemOpen
  \bibfield  {author} {\bibinfo {author} {\bibfnamefont {T.}~\bibnamefont {Graham}}, \bibinfo {author} {\bibfnamefont {Y.}~\bibnamefont {Song}}, \bibinfo {author} {\bibfnamefont {J.}~\bibnamefont {Scott}},  \emph {et~al.},\ }\href {\doibase 10.1038/s41586-022-04603-6} {\bibfield  {journal} {\bibinfo  {journal} {Nature}\ }\textbf {\bibinfo {volume} {604}},\ \bibinfo {pages} {457–462} (\bibinfo {year} {2022})}\BibitemShut {NoStop}%
\bibitem [{\citenamefont {Evered}\ \emph {et~al.}(2023)\citenamefont {Evered}, \citenamefont {Bluvstein}, \citenamefont {D.} \emph {et~al.}}]{Evered23}%
  \BibitemOpen
  \bibfield  {author} {\bibinfo {author} {\bibfnamefont {S.}~\bibnamefont {Evered}}, \bibinfo {author} {\bibnamefont {Bluvstein}}, \bibinfo {author} {\bibfnamefont {M.}~\bibnamefont {D.}, \bibfnamefont {Kalinowski}},  \emph {et~al.},\ }\href {\doibase 10.1038/s41586-023-06481-y} {\bibfield  {journal} {\bibinfo  {journal} {Nature}\ }\textbf {\bibinfo {volume} {622}},\ \bibinfo {pages} {268–272} (\bibinfo {year} {2023})}\BibitemShut {NoStop}%
\bibitem [{\citenamefont {Einstein}\ \emph {et~al.}(1935)\citenamefont {Einstein}, \citenamefont {Podolsky},\ and\ \citenamefont {Rosen}}]{Einstein35}%
  \BibitemOpen
  \bibfield  {author} {\bibinfo {author} {\bibfnamefont {A.}~\bibnamefont {Einstein}}, \bibinfo {author} {\bibfnamefont {B.}~\bibnamefont {Podolsky}}, \ and\ \bibinfo {author} {\bibfnamefont {N.}~\bibnamefont {Rosen}},\ }\href {\doibase 10.1103/PhysRev.47.777} {\bibfield  {journal} {\bibinfo  {journal} {Phys. Rev.}\ }\textbf {\bibinfo {volume} {47}},\ \bibinfo {pages} {777} (\bibinfo {year} {1935})}\BibitemShut {NoStop}%
\bibitem [{\citenamefont {Ishikawa}\ \emph {et~al.}(2023)\citenamefont {Ishikawa}, \citenamefont {Prince},\ and\ \citenamefont {Ueda}}]{Ishikawa23}%
  \BibitemOpen
  \bibfield  {author} {\bibinfo {author} {\bibfnamefont {K.~L.}\ \bibnamefont {Ishikawa}}, \bibinfo {author} {\bibfnamefont {K.~C.}\ \bibnamefont {Prince}}, \ and\ \bibinfo {author} {\bibfnamefont {K.}~\bibnamefont {Ueda}},\ }\href {\doibase 10.1021/acs.jpca.3c06781} {\bibfield  {journal} {\bibinfo  {journal} {J. Phys. Chem. A}\ }\textbf {\bibinfo {volume} {127}},\ \bibinfo {pages} {10638} (\bibinfo {year} {2023})}\BibitemShut {NoStop}%
\bibitem [{\citenamefont {Laurell}\ \emph {et~al.}(2025)\citenamefont {Laurell}, \citenamefont {Luo}, \citenamefont {Weissenbilder} \emph {et~al.}}]{Laurell25}%
  \BibitemOpen
  \bibfield  {author} {\bibinfo {author} {\bibfnamefont {H.}~\bibnamefont {Laurell}}, \bibinfo {author} {\bibfnamefont {S.}~\bibnamefont {Luo}}, \bibinfo {author} {\bibfnamefont {R.}~\bibnamefont {Weissenbilder}},  \emph {et~al.},\ }\href {\doibase 0.1038/s41566-024-01607-8} {\bibfield  {journal} {\bibinfo  {journal} {Nat. Photon.}\ }\textbf {\bibinfo {volume} {19}},\ \bibinfo {pages} {352–357} (\bibinfo {year} {2025})}\BibitemShut {NoStop}%
\bibitem [{\citenamefont {Berkane}\ \emph {et~al.}(2025)\citenamefont {Berkane}, \citenamefont {Ta\"{\i}eb}, \citenamefont {Granveau}, \citenamefont {Sali\`eres}, \citenamefont {Bourassin-Bouchet}, \citenamefont {L\'ev\^eque},\ and\ \citenamefont {Caillat}}]{Berkane25}%
  \BibitemOpen
  \bibfield  {author} {\bibinfo {author} {\bibfnamefont {M.}~\bibnamefont {Berkane}}, \bibinfo {author} {\bibfnamefont {R.}~\bibnamefont {Ta\"{\i}eb}}, \bibinfo {author} {\bibfnamefont {G.}~\bibnamefont {Granveau}}, \bibinfo {author} {\bibfnamefont {P.}~\bibnamefont {Sali\`eres}}, \bibinfo {author} {\bibfnamefont {C.}~\bibnamefont {Bourassin-Bouchet}}, \bibinfo {author} {\bibfnamefont {C.}~\bibnamefont {L\'ev\^eque}}, \ and\ \bibinfo {author} {\bibfnamefont {J.}~\bibnamefont {Caillat}},\ }\href {\doibase 10.1103/PhysRevA.111.L041101} {\bibfield  {journal} {\bibinfo  {journal} {Phys. Rev. A}\ }\textbf {\bibinfo {volume} {111}},\ \bibinfo {pages} {L041101} (\bibinfo {year} {2025})}\BibitemShut {NoStop}%
\bibitem [{\citenamefont {Shen}\ \emph {et~al.}(2025)\citenamefont {Shen}, \citenamefont {Mao}, \citenamefont {Zhang}, \citenamefont {Li}, \citenamefont {Sato}, \citenamefont {Ishikawa},\ and\ \citenamefont {He}}]{Shen25}%
  \BibitemOpen
  \bibfield  {author} {\bibinfo {author} {\bibfnamefont {B.-R.}\ \bibnamefont {Shen}}, \bibinfo {author} {\bibfnamefont {Y.-J.}\ \bibnamefont {Mao}}, \bibinfo {author} {\bibfnamefont {Z.-H.}\ \bibnamefont {Zhang}}, \bibinfo {author} {\bibfnamefont {Y.}~\bibnamefont {Li}}, \bibinfo {author} {\bibfnamefont {T.}~\bibnamefont {Sato}}, \bibinfo {author} {\bibfnamefont {K.~L.}\ \bibnamefont {Ishikawa}}, \ and\ \bibinfo {author} {\bibfnamefont {F.}~\bibnamefont {He}},\ }\href {\doibase 10.1103/c338-8n6w} {\bibfield  {journal} {\bibinfo  {journal} {Phys. Rev. A}\ }\textbf {\bibinfo {volume} {111}},\ \bibinfo {pages} {063113} (\bibinfo {year} {2025})}\BibitemShut {NoStop}%
\bibitem [{\citenamefont {Nandi}\ \emph {et~al.}(2024)\citenamefont {Nandi}, \citenamefont {Stenquist}, \citenamefont {Papoulia} \emph {et~al.}}]{Nandi24}%
  \BibitemOpen
  \bibfield  {author} {\bibinfo {author} {\bibfnamefont {S.}~\bibnamefont {Nandi}}, \bibinfo {author} {\bibfnamefont {A.}~\bibnamefont {Stenquist}}, \bibinfo {author} {\bibfnamefont {A.}~\bibnamefont {Papoulia}},  \emph {et~al.},\ }\href {\doibase 10.1126/sciadv.ado0668} {\bibfield  {journal} {\bibinfo  {journal} {Science Advances}\ }\textbf {\bibinfo {volume} {10}},\ \bibinfo {pages} {eado0668} (\bibinfo {year} {2024})}\BibitemShut {NoStop}%
\bibitem [{\citenamefont {McNeil}\ and\ \citenamefont {Thompson}(2010)}]{McNeil10}%
  \BibitemOpen
  \bibfield  {author} {\bibinfo {author} {\bibfnamefont {B.}~\bibnamefont {McNeil}}\ and\ \bibinfo {author} {\bibfnamefont {N.}~\bibnamefont {Thompson}},\ }\href {\doibase https://doi.org/10.1038/nphoton.2010.239} {\bibfield  {journal} {\bibinfo  {journal} {Nature Photon}\ }\textbf {\bibinfo {volume} {4}},\ \bibinfo {pages} {814–821} (\bibinfo {year} {2010})}\BibitemShut {NoStop}%
\bibitem [{\citenamefont {Grobe}\ and\ \citenamefont {Eberly}(1993)}]{Grobe93}%
  \BibitemOpen
  \bibfield  {author} {\bibinfo {author} {\bibfnamefont {R.}~\bibnamefont {Grobe}}\ and\ \bibinfo {author} {\bibfnamefont {J.~H.}\ \bibnamefont {Eberly}},\ }\href {\doibase 10.1103/PhysRevA.48.623} {\bibfield  {journal} {\bibinfo  {journal} {Phys. Rev. A}\ }\textbf {\bibinfo {volume} {48}},\ \bibinfo {pages} {623} (\bibinfo {year} {1993})}\BibitemShut {NoStop}%
\bibitem [{\citenamefont {Girju}\ \emph {et~al.}(2007)\citenamefont {Girju}, \citenamefont {Hristov}, \citenamefont {Kidun},\ and\ \citenamefont {Bauer}}]{Girju07}%
  \BibitemOpen
  \bibfield  {author} {\bibinfo {author} {\bibfnamefont {M.~G.}\ \bibnamefont {Girju}}, \bibinfo {author} {\bibfnamefont {K.}~\bibnamefont {Hristov}}, \bibinfo {author} {\bibfnamefont {O.}~\bibnamefont {Kidun}}, \ and\ \bibinfo {author} {\bibfnamefont {D.}~\bibnamefont {Bauer}},\ }\href {\doibase 10.1088/0953-4075/40/21/004} {\bibfield  {journal} {\bibinfo  {journal} {Journal of Physics B: Atomic, Molecular and Optical Physics}\ }\textbf {\bibinfo {volume} {40}},\ \bibinfo {pages} {4165} (\bibinfo {year} {2007})}\BibitemShut {NoStop}%
\bibitem [{\citenamefont {Rohringer}\ and\ \citenamefont {Santra}(2008)}]{Rohringer08}%
  \BibitemOpen
  \bibfield  {author} {\bibinfo {author} {\bibfnamefont {N.}~\bibnamefont {Rohringer}}\ and\ \bibinfo {author} {\bibfnamefont {R.}~\bibnamefont {Santra}},\ }\href {\doibase 10.1103/PhysRevA.77.053404} {\bibfield  {journal} {\bibinfo  {journal} {Phys. Rev. A}\ }\textbf {\bibinfo {volume} {77}},\ \bibinfo {pages} {053404} (\bibinfo {year} {2008})}\BibitemShut {NoStop}%
\bibitem [{\citenamefont {Demekhin}\ and\ \citenamefont {Cederbaum}(2012{\natexlab{a}})}]{Demekhin12_a}%
  \BibitemOpen
  \bibfield  {author} {\bibinfo {author} {\bibfnamefont {P.~V.}\ \bibnamefont {Demekhin}}\ and\ \bibinfo {author} {\bibfnamefont {L.~S.}\ \bibnamefont {Cederbaum}},\ }\href {\doibase 10.1103/PhysRevLett.108.253001} {\bibfield  {journal} {\bibinfo  {journal} {Phys. Rev. Lett.}\ }\textbf {\bibinfo {volume} {108}},\ \bibinfo {pages} {253001} (\bibinfo {year} {2012}{\natexlab{a}})}\BibitemShut {NoStop}%
\bibitem [{\citenamefont {Demekhin}\ and\ \citenamefont {Cederbaum}(2012{\natexlab{b}})}]{Demekhin12_b}%
  \BibitemOpen
  \bibfield  {author} {\bibinfo {author} {\bibfnamefont {P.~V.}\ \bibnamefont {Demekhin}}\ and\ \bibinfo {author} {\bibfnamefont {L.~S.}\ \bibnamefont {Cederbaum}},\ }\href {\doibase 10.1103/PhysRevA.86.063412} {\bibfield  {journal} {\bibinfo  {journal} {Phys. Rev. A}\ }\textbf {\bibinfo {volume} {86}},\ \bibinfo {pages} {063412} (\bibinfo {year} {2012}{\natexlab{b}})}\BibitemShut {NoStop}%
\bibitem [{\citenamefont {Zhang}\ and\ \citenamefont {Rohringer}(2014)}]{Zhang14}%
  \BibitemOpen
  \bibfield  {author} {\bibinfo {author} {\bibfnamefont {S.~B.}\ \bibnamefont {Zhang}}\ and\ \bibinfo {author} {\bibfnamefont {N.}~\bibnamefont {Rohringer}},\ }\href {\doibase 10.1103/PhysRevA.89.013407} {\bibfield  {journal} {\bibinfo  {journal} {Phys. Rev. A}\ }\textbf {\bibinfo {volume} {89}},\ \bibinfo {pages} {013407} (\bibinfo {year} {2014})}\BibitemShut {NoStop}%
\bibitem [{\citenamefont {Deng}\ \emph {et~al.}(2025)\citenamefont {Deng}, \citenamefont {Zhang}, \citenamefont {Zhou},\ and\ \citenamefont {Lu}}]{Deng25}%
  \BibitemOpen
  \bibfield  {author} {\bibinfo {author} {\bibfnamefont {Y.}~\bibnamefont {Deng}}, \bibinfo {author} {\bibfnamefont {X.}~\bibnamefont {Zhang}}, \bibinfo {author} {\bibfnamefont {Y.}~\bibnamefont {Zhou}}, \ and\ \bibinfo {author} {\bibfnamefont {P.}~\bibnamefont {Lu}},\ }\href {\doibase 10.1103/PhysRevA.111.043110} {\bibfield  {journal} {\bibinfo  {journal} {Phys. Rev. A}\ }\textbf {\bibinfo {volume} {111}},\ \bibinfo {pages} {043110} (\bibinfo {year} {2025})}\BibitemShut {NoStop}%
\bibitem [{\citenamefont {Wan}\ \emph {et~al.}(2025)\citenamefont {Wan}, \citenamefont {Liu}, \citenamefont {Xie}, \citenamefont {Meng},\ and\ \citenamefont {Jiang}}]{Wan25}%
  \BibitemOpen
  \bibfield  {author} {\bibinfo {author} {\bibfnamefont {W.-Q.}\ \bibnamefont {Wan}}, \bibinfo {author} {\bibfnamefont {G.-Y.}\ \bibnamefont {Liu}}, \bibinfo {author} {\bibfnamefont {M.-F.}\ \bibnamefont {Xie}}, \bibinfo {author} {\bibfnamefont {Q.-B.}\ \bibnamefont {Meng}}, \ and\ \bibinfo {author} {\bibfnamefont {W.-C.}\ \bibnamefont {Jiang}},\ }\href {\doibase 10.1103/PhysRevA.111.033117} {\bibfield  {journal} {\bibinfo  {journal} {Phys. Rev. A}\ }\textbf {\bibinfo {volume} {111}},\ \bibinfo {pages} {033117} (\bibinfo {year} {2025})}\BibitemShut {NoStop}%
\bibitem [{\citenamefont {Nandi}\ \emph {et~al.}(2022)\citenamefont {Nandi}, \citenamefont {Olofsson}, \citenamefont {Bertolino} \emph {et~al.}}]{Nandi22}%
  \BibitemOpen
  \bibfield  {author} {\bibinfo {author} {\bibfnamefont {S.}~\bibnamefont {Nandi}}, \bibinfo {author} {\bibfnamefont {E.}~\bibnamefont {Olofsson}}, \bibinfo {author} {\bibfnamefont {M.}~\bibnamefont {Bertolino}},  \emph {et~al.},\ }\href {\doibase https://doi.org/10.1038/s41586-022-04948-y} {\bibfield  {journal} {\bibinfo  {journal} {Nature}\ }\textbf {\bibinfo {volume} {608}},\ \bibinfo {pages} {488–493} (\bibinfo {year} {2022})}\BibitemShut {NoStop}%
\bibitem [{\citenamefont {Stenquist}\ and\ \citenamefont {Dahlstr\"om}(2025)}]{Stenquist25}%
  \BibitemOpen
  \bibfield  {author} {\bibinfo {author} {\bibfnamefont {A.}~\bibnamefont {Stenquist}}\ and\ \bibinfo {author} {\bibfnamefont {J.~M.}\ \bibnamefont {Dahlstr\"om}},\ }\href {\doibase 10.1103/PhysRevResearch.7.013270} {\bibfield  {journal} {\bibinfo  {journal} {Phys. Rev. Res.}\ }\textbf {\bibinfo {volume} {7}},\ \bibinfo {pages} {013270} (\bibinfo {year} {2025})}\BibitemShut {NoStop}%
\bibitem [{\citenamefont {Yu}\ and\ \citenamefont {Madsen}(2018)}]{Yu18}%
  \BibitemOpen
  \bibfield  {author} {\bibinfo {author} {\bibfnamefont {C.}~\bibnamefont {Yu}}\ and\ \bibinfo {author} {\bibfnamefont {L.~B.}\ \bibnamefont {Madsen}},\ }\href {\doibase 10.1103/PhysRevA.98.033404} {\bibfield  {journal} {\bibinfo  {journal} {Phys. Rev. A}\ }\textbf {\bibinfo {volume} {98}},\ \bibinfo {pages} {033404} (\bibinfo {year} {2018})}\BibitemShut {NoStop}%
\bibitem [{\citenamefont {Tao}\ and\ \citenamefont {Scrinzi}(2012)}]{Tao12}%
  \BibitemOpen
  \bibfield  {author} {\bibinfo {author} {\bibfnamefont {L.}~\bibnamefont {Tao}}\ and\ \bibinfo {author} {\bibfnamefont {A.}~\bibnamefont {Scrinzi}},\ }\href {\doibase 10.1088/1367-2630/14/1/013021} {\bibfield  {journal} {\bibinfo  {journal} {New Journal of Physics}\ }\textbf {\bibinfo {volume} {14}},\ \bibinfo {pages} {013021} (\bibinfo {year} {2012})}\BibitemShut {NoStop}%
\bibitem [{\citenamefont {Scrinzi}(2012)}]{Scrinzi12}%
  \BibitemOpen
  \bibfield  {author} {\bibinfo {author} {\bibfnamefont {A.}~\bibnamefont {Scrinzi}},\ }\href {\doibase 10.1088/1367-2630/14/8/085008} {\bibfield  {journal} {\bibinfo  {journal} {New Journal of Physics}\ }\textbf {\bibinfo {volume} {14}},\ \bibinfo {pages} {085008} (\bibinfo {year} {2012})}\BibitemShut {NoStop}%
\bibitem [{\citenamefont {Ramsey}(1950)}]{Ramsey50}%
  \BibitemOpen
  \bibfield  {author} {\bibinfo {author} {\bibfnamefont {N.~F.}\ \bibnamefont {Ramsey}},\ }\href {\doibase 10.1103/PhysRev.78.695} {\bibfield  {journal} {\bibinfo  {journal} {Phys. Rev.}\ }\textbf {\bibinfo {volume} {78}},\ \bibinfo {pages} {695} (\bibinfo {year} {1950})}\BibitemShut {NoStop}%
\bibitem [{\citenamefont {Dittel}\ \emph {et~al.}(2021)\citenamefont {Dittel}, \citenamefont {Dufour}, \citenamefont {Weihs},\ and\ \citenamefont {Buchleitner}}]{Dittel21}%
  \BibitemOpen
  \bibfield  {author} {\bibinfo {author} {\bibfnamefont {C.}~\bibnamefont {Dittel}}, \bibinfo {author} {\bibfnamefont {G.}~\bibnamefont {Dufour}}, \bibinfo {author} {\bibfnamefont {G.}~\bibnamefont {Weihs}}, \ and\ \bibinfo {author} {\bibfnamefont {A.}~\bibnamefont {Buchleitner}},\ }\href {\doibase 10.1103/PhysRevX.11.031041} {\bibfield  {journal} {\bibinfo  {journal} {Phys. Rev. X}\ }\textbf {\bibinfo {volume} {11}},\ \bibinfo {pages} {031041} (\bibinfo {year} {2021})}\BibitemShut {NoStop}%
\bibitem [{\citenamefont {Blanes}\ and\ \citenamefont {Moan}(2002)}]{Blanes02}%
  \BibitemOpen
  \bibfield  {author} {\bibinfo {author} {\bibfnamefont {S.}~\bibnamefont {Blanes}}\ and\ \bibinfo {author} {\bibfnamefont {P.}~\bibnamefont {Moan}},\ }\href {\doibase 10.1016/s0377-0427(01)00492-7} {\bibfield  {journal} {\bibinfo  {journal} {Journal of Computational and Applied Mathematics}\ }\textbf {\bibinfo {volume} {142}},\ \bibinfo {pages} {313} (\bibinfo {year} {2002})}\BibitemShut {NoStop}%
\bibitem [{\citenamefont {McLachlan}(2022)}]{McLachlan22}%
  \BibitemOpen
  \bibfield  {author} {\bibinfo {author} {\bibfnamefont {R.~I.}\ \bibnamefont {McLachlan}},\ }\href {\doibase 10.4208/cicp.oa-2021-0154} {\bibfield  {journal} {\bibinfo  {journal} {Communications in Computational Physics}\ }\textbf {\bibinfo {volume} {31}},\ \bibinfo {pages} {987} (\bibinfo {year} {2022})}\BibitemShut {NoStop}%
\end{thebibliography}%

\end{document}